\documentstyle[twoside,fleqn,epsf,espcrc2]{article}

\def\d#1/d#2{ {\partial #1\over\partial #2} }

\def\tr{\hbox{tr}}

\def\beq{\begin{equation}}
\def\eeq{\end{equation}}
\def\beqa{\begin{eqnarray*}}
\def\eeqa{\end{eqnarray*}}
\def\beqs{\begin{eqnarray}}
\def\eeqs{\end{eqnarray}}
\def\n{\global\advance \eqnumber by 1\eqno(\the\eqnumber)}
\def\puteqno{\global\advance \eqnumber by 1 (\the\eqnumber)}

\newcommand{\npb}[3]{~{Nucl. Phys.}~{B#1}~{(19#2)}~{#3.}} 
\newcommand{\prd}[3]{~{Phys. Rev.}~{D#1}~{(19#2)}~{#3.}}

\def\bn{{\bf n}}
\title{Monopole-antimonopole condensation in the interpolating
Georgi-Glashow model}
\author{Chulwoo Jung \address{Department of Physics, Columbia University, New
York, NY 10027 U.S.A.}
        \thanks{This work was done in collaboration with Robert D.
Mawhinney. It was partially supported by the US Department of Energy.
} }
\begin{document}
\begin{abstract}
We study the three dimensional Georgi-Glashow model (which
interpolates smoothly between pure U(1) and SU(2) limits) using a
constrained cooling which preserves 't Hooft-Polyakov monopoles. 
We find that the monopole-antimonopole condensation gives an area law 
for the Wilson loops. The monopole contribution to the string tension
is close to the Monte Carlo value in the intermediate region.
\end{abstract}
\maketitle
In spite of considerable evidence for confinement at weak coupling from
numerical simulations, an analytic understanding of the long-range
properties of non-Abelian gauge theories has been quite limited. 
In this note, we investigate the role of the 't Hooft-Polyakov monopoles
 in the non-perturbative physics of the 3 dimensional Georgi-Glashow 
model($GG_3$).  It was shown by
Polyakov\cite{pol} that the condensation of these monopoles is
responsible for confinement in the region where monopoles
are far apart and the distribution of the monopoles is determined by
Coulomb interaction between monopoles. 
Cooling (systematic reduction of the action)\cite{dunmaw},
when applied to the Monte Carlo lattices in the intermediate region of
$GG_3$, yields collections of 
(anti)monopoles which are stable solutions on the lattice. 
However, the distribution of monopoles in these
configurations may have little bearing on
the monopole distribution in the original configurations 
since cooling can change the monopole distribution.

In this work, the role of the 't Hooft-Polyakov monopoles present 
in the Monte Carlo(MC) lattices is investigated 
using a constrained cooling procedure we
refer as monopole preserving cooling (MP cooling in short) which
preserves the number and the location of the monopoles 
while cooling.
A brief description of MP
cooling and the comparison between MP cooling and the unconstrained
cooling(`cooling' in short) is given in section 1. We discuss the
behavior of the Creutz ratios (both fundamental and  adjoint)  of MP cooled lattices in section 2.

\section{MP cooling}
\label{section:cooling}
For the numerical simulation, we used a discretized version of
the Georgi-Glashow Lagrangian with the radially frozen approximation.
($\vec \Phi_\bn$ is a real 3-vector, $\vec{\Phi}_{\bf n} \cdot \vec{\Phi}_{\bf n} = 1$)
\begin{eqnarray*}
S(\beta_g,\beta_h)= \beta_g S_g + \beta_h S_h \equiv \beta_g\sum_{P}(1- \frac12 \tr({U_P})) \\
\qquad\ + \beta_h\sum_{{\bf n} \mu}(1-\frac12 \tr( \vec{ \Phi}_{\bf n} \cdot \vec{\sigma}
U_{{\bf n}\mu} \vec\Phi_{{\bf n}+\hat{\mu}} \cdot \vec{\sigma} U_{{\bf n}\mu}^{\dagger}))
\end{eqnarray*}
This model describes a U(1) theory as 
$\beta_h \rightarrow \infty$ and the pure SU(2) theory as 
as $\beta_h \rightarrow 0$. 

MP cooling  on the MC configurations is implemented as follows.
After gauge transforming the Higgs fields to 
$\vec\Phi = \hat{z}$, the projected 
U(1) field is given by $\theta_{\bn \mu} = \arg((U_{\bn \mu})_{11})
(-\pi < \theta < \pi)$. Now the monopoles are located on the dual
lattice points by applying a standard procedure devised by DeGrand \&
Toussaint\cite{mpcount} on $\theta_{\bn \mu}$.
A new gauge link which lowers the action is  
accepted only if the monopole numbers for the 4 cubes 
surrounding the gauge link remains the same. 

\small
\begin{table*}[hbt]
\begin{tabular*}{\textwidth}{@{\extracolsep{\fill}}cccccc} \hline
$\beta_g$&3.0 & 6.5&7.6 & 9.4 & 9.9 \\
$\beta_h$&3.0&1.0&0.87 & 0.5 & 0.1 \\ \hline
$S_g$/plaquette&0.26341(10) &0.15472(3)&0.13422(3)&0.1094(5)&0.1038(5)\\
$S_h$/link&0.22705(6)&0.5068(3) &0.5962(3)&0.8200(33)&0.9667(28)\\
MP $S_g$/plaquette&0.00174(7)&0.00144(2)&0.002086(7)&0.00114(3)&0.00012(2)\\
MP $S_h$/link&0.000196(8)&0.0434(4)&0.1181(3)&0.4003(59)&0.8086(33)\\
Monopole density&0.01494(4) &0.0268(2) &0.0646(2)&0.1984(2) & 0.2776(2)\\ 
MC String tension&0.0288(19)&0.0206(19) &0.0323(9) &0.0271(12) 
&0.0241(5) \\
MP Creutz ratio&0.0216(21)&0.0119(20)&0.0268(18)&0.0157(7)&\\
\hline
\end{tabular*}
\caption{The summary of measured quantities.  All the simulations are
done on a $24^3$ lattice. The quoted value of the MC string
tension is the constant fit to the Creutz ratio $C(R,R)$ for $R=8$ to $R=12$.
MP Creutz ratio is $C(R,R)$ at $R=12$. MP Creutz ratio may not be close
 to the asymptotic value due to the slow convergence as you see in
Fig.~\ref{fig:cr}.}

\label{table:actmpl}
\end{table*}
\normalsize
Table~\ref{table:actmpl} shows the action density and the monopole
density for $24^3$ lattices. We used the couplings ($\beta_g,\beta_h$) 
obtained by Duncan and
Mawhinney, which keep the string tension qualitatively the 
same\cite{dunmaw}. At $\beta_g=3.0,\beta_h=3.0$, the
MP cooled configuration is essentially U(1)($S_h \sim 0$) since the nonabelian
cores of the monopoles are small. As we approach the SU(2) limit,
the monopole density increases since the action of individual
monopoles 
 gets smaller. 

Table~\ref{table:mpl1} shows the evolution of the 
monopole number and the action density of a single
lattice near the U(1) limit($\beta_g=3.0,\beta_g=3.0$) and near the
SU(2) limit ($\beta_g=9.9,\beta_h=0.1$)
under cooling and MP cooling. Cooling induces the annihilation of close
monopole-antimonopole pairs, resulting in a lower action 
than MP cooling.
\small
\begin{table}[hbt]
\vspace{-3mm}
\begin{center}
\begin{tabular}{c|ccc|cc}\hline
\multicolumn{6}{c}{$\beta_g=3.0\;\;\beta_h=3.0$}\\ \hline
&\multicolumn{3}{c|}{cooling}&\multicolumn{2}{c}{MP cooling}\\ \hline 
$N$&mpl&$S_g$&$S_h$&$S_g$&$S_h$\\ \hline
0&24&0.2629&0.2284&0.2629&0.2284\\
2&16&8.82e-2&9.62e-2&8.82e-2&9.61e-2\\
10&12&1.58e-3&1.69e-4&2.08e-2&4.83e-4\\
50&12&1.26e-3&1.1e-10&1.75e-3&3.41e-4\\ \hline
\multicolumn{6}{c}{$\beta_g=9.9\;\;\beta_h=0.1$}\\ \hline
&\multicolumn{3}{c|}{cooling}&\multicolumn{2}{c}{MP cooling}\\ \hline 
$N$&mpl&$S_g$&$S_h$&$S_g$&$S_h$\\ \hline
0&3872&0.1041 &0.9696 &0.1041&0.9696\\ 
10&3446&3.35e-4&0.9338&5.41e-4&0.9389\\
50&2534&9.95e-5&0.7728&1.65e-4&0.8111\\
100&1606&6.17e-5&0.5893&1.24e-4&0.6979\\
200&616&4.03e-5&0.3057&1.04e-4&0.5977\\
300&210&3.63e-5&0.1434&9.50e-5&0.5663\\ \hline
\end{tabular}
\end{center}
\caption{The evolution of the monopole number and the action density
under cooling and MP cooling. For MP cooling, the monopole number is
fixed. cooling. N is the number of cooling steps.} 
\label{table:mpl1}
\end{table}
\normalsize

\begin{figure}[hbt]
\epsfxsize=3.0in
\epsfbox[-50 0 356 250 ]{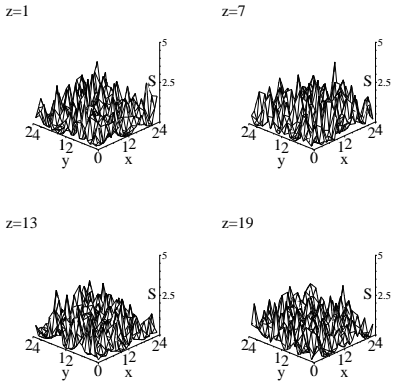}
\vspace{-10mm}
\caption{Action density plot of a MP cooled lattice for
$\beta_g=9.4\;\;\beta_h=0.5$.}
\label{fig:act1}
\end{figure}
\vspace{-5mm}
Fig.~\ref{fig:act1} shows the action density plot of a MP cooled lattice
near the SU(2) limit ($\beta_g=9.4,\beta_h=0.5$). In contrast to
the smooth action density we see for cooled lattices, MP cooled
lattices 
show many small but sharp peaks and seems to lack any global structure
which might develops into the smooth peaks when we approach the
SU(2) limit. 
\section{Creutz ratios}
\begin{figure}
\epsfxsize=2.7in
\epsfbox[18 43 572 428]{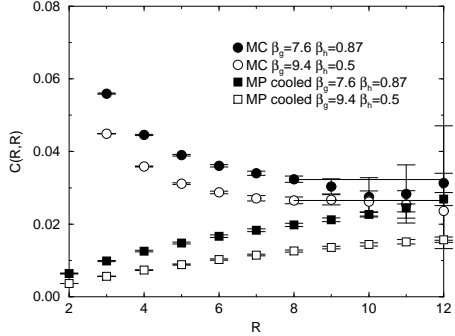}
\vspace{-1cm}
\caption{Creutz ratios for the MC and MP cooled lattices. Straight
lines indicates the fitted value of the string tension for the MC
lattices.}
\label{fig:cr}
\end{figure}

Fig.~\ref{fig:cr} shows the behavior of the 
fundamental Creutz ratios for the MC
and  the MP cooled lattices in the intermediate
region ($\beta_g=7.6,\beta_h=0.87$) and near the SU(2)
limit ($\beta_g=9.4,\beta_h=0.5$). 
While the presence of the Coulomb interaction makes the fundamental 
Creutz ratio for the MC lattices approach the 
asymptotic value from above, the Creutz ratio for MP cooled lattices
approaches from below, similar to the behavior of the Creutz ratio of
the monopole contribution in the Villain U(1) model\cite{dunmaw}. 
From near the U(1) limit to the intermediate region, the MP Creutz
ratios 
approach the MC Creutz ratios at a large $R$ and a
plateau begins  to emerge.
As we move closer to the SU(2) limit where the nonabelian cores of 
the monopoles begin to overlap, 
the area law for the Wilson loops persists although the Creutz ratio
increases slower. It is not clear if the MP Creutz ratio converges to a
smaller value or the lattice size is too small compared to the length
scale at which it converges.

\begin{figure}[hbt]
\epsfxsize=2.7in
\epsfbox[18 43 572 428]{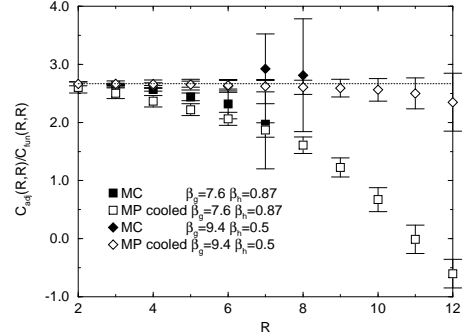}
\vspace{-1cm}
\caption{Ratios of the fundamental and the adjoint Creutz ratio of
the MC and the MP cooled lattice. The dotted line indicates the Casimir ratio(8/3). The
ratio is zero at a large $R$ for both the MC and the MP cooled lattices at
$\beta_g=3.0,\beta_h=3.0$.}
\label{fig:ratio}
\end{figure}

Fig.~\ref{fig:ratio} shows the ratios of the fundamental and the
adjoint Creutz ratio for the MC and the MP cooled lattices. 
The ratio is asymptotically  0 for a U(1) theory and 
is 8/3 for pure SU(2) gauge theory until the onset of the color 
screening (Casimir scaling)\cite{greensite}. In the intermediate
region, the ratios for both the MC and MP cooled lattices show
the transition between two values. 
Although the individual (fundamental, adjoint) Creutz ratios are quite 
different, the ratios of  two Creutz ratios
for MP cooled lattices show the same features as the MC lattices.
\section{Discussion}
The 't Hooft-Polyakov monopole condensation in $GG_3$ gives an 
area law for the Wilson loops 
although the string tension appears to be less than the MC value,
especially near the SU(2) limit. 
The non-vanishing Higgs action and the Casimir scaling of the MP cooled 
lattices near the SU(2) limit suggests the interactions between these monopoles are no 
longer abelian.  More simulation with a larger volume is needed to 
see whether the smaller Creutz ratios for MP cooled lattices are due simply 
to the small volume or a change in the confinement mechanism in
this region.

\end{document}